\documentclass[10pt]{article}
\usepackage{hyperref}
\usepackage{geometry}
\usepackage{graphicx}
\usepackage{slashed}
\usepackage{cite}
\usepackage{amssymb}
\usepackage{amsmath}
\usepackage{dsfont}
\usepackage{color}
\usepackage{indentfirst}
\usepackage{multirow}
\usepackage{subfigure}
\usepackage{float}
\usepackage{makecell}
\allowdisplaybreaks[4]

\begin{document}

\title{\huge\bf Studies on quark-mass dependence of the $N^*(920)$ pole from $\pi N$ $\chi$PT amplitudes}

\author{Xu Wang\textsuperscript{1}\footnote{wangxu0604@stu.pku.edu.cn}, Kai-Ge Kang\textsuperscript{1,2}\footnote{kaige@alumni.pku.edu.cn;
 part of this work is done while the author was visiting SCU.}, 
Qu-Zhi Li\textsuperscript{2}\footnote{liqz@scu.edu.cn (corresponding author)},
Zhiguang Xiao\textsuperscript{2}\footnote{xiaozg@scu.edu.cn (corresponding author)}
and Han-Qing Zheng\textsuperscript{2}\footnote{zhenghq@scu.edu.cn}\\
\small \textsuperscript{1} School of Physics, Peking University, Beijing 100871, China\\
\small \textsuperscript{2} Institute of Particle and Nuclear Physics, 
Sichuan University, Chengdu, Sichuan 610065, China
}
\maketitle

\begin{abstract}
 The quark-mass dependence of the $N^*(920)$ pole is analyzed using $K$-matrix method, with the $\pi N$ scattering amplitude calculated up to $O(p^3)$  order in chiral perturbation theory. As the quark mass increases, the $N^*(920)$ pole gradually approaches the real axis in the complex $w$-plane (where $w=\sqrt{s}$). Eventually, in the $O(p^2)$ case, it crosses the $u$-cut on the real axis and enters the adjacent Riemann sheet when the pion mass reaches $526~{\rm MeV}$. At order $O(p^3)$, the rate at which it approaches the real axis slows down; however, we argue that it will ultimately cross the $u$-cut and enter the adjacent Riemann sheet as well. Additionally, the trajectory of the \(N^*(920)\) pole is in qualitative agreement with the results from the linear $\sigma$ model calculation.
\end{abstract}

\section{Introduction}\label{sec:i}
    The study of pion-nucleon scattering has a history spanning over sixty years. It is therefore surprising that the pole structure of the sub-threshold $\pi-N$ scattering amplitude, particularly in the $S_{11}$ channel, has only been clarified very recently. Two key findings have emerged: first, as demonstrated in Ref.~\cite{Li:2021oou}, partial-wave amplitudes (PWAs) indeed contain poles --- specifically, virtual states --- located on the real axis below threshold on the second Riemann sheet (RSII). Second, a novel resonance pole has been identified in the $S_{11}$ channel through various unitarized approaches, including the product  representation~\cite{Zheng:2003rw,Wang:2017agd,Wang:2018nwi}, the $K$-matrix fit~\cite{ctp20may}, and the $N/D$ method~\cite{cpc22quz}. The existence of this resonance has been further confirmed by the model-independent Roy-Steiner equation formalism~\cite{Cao:2022zhn,Hoferichter:2023mgy}, which respects  analyticity, unitarity, and crossing symmetry of the $S$-matrix. This sub-threshold pole, located at $\sqrt{s}= (918 \pm 3) - i(163 \pm 9)~\mathrm{MeV}$, has been designated as $N^*(920)$.

    Meanwhile, understanding the quark-mass dependence of resonance poles is crucial, which offers a unique perspective on strong interaction physics. Lattice QCD provides a first-principle, non-perturbative framework to investigate how hadron states depend on the quark mass. However, parameterizations of infinite-volume PWAs can introduce model dependence when fitting finite-volume spectra using the L\"uscher formula~\cite{Luscher:1990ux} and its generalizations~\cite{Rummukainen:1995vs, Fu:2011xz, Leskovec:2012gb}. Ref.~\cite{Cao:2023ntr} demonstrated a model-independent approach to interpreting lattice data via the generalized Roy equation, revealing that the $\sigma$ meson becomes a bound state, with a new resonance emerging, at $m_\pi\simeq 391\mathrm{MeV}$. Similar studies have also been completed for $\pi K$ scattering, as detailed in Refs.~\cite{Cao:2024zuy,Cao:2025hqm}. Subsequently, the trajectory of the $\sigma$ with varying $m_\pi$ was illustrated within the $O(N)$ linear $\sigma$ model (L$\sigma$M)~\cite{Lyu:2024elz,Lyu:2024lzr}.

    The first attempt to trace the trajectory of the $N^*(920)$ with varying pion masses was conducted within the L$\sigma$M with nucleons~\cite{Li:2025fvg}. In that renormalizable model, the authors simultaneously computed the trajectories of both the $\sigma$ and the $N^*(920)$ using several unitarization methods at the one-loop level. The trajectory of the $\sigma$ was found to be consistent with previous results, while that of the $N^*(920)$ was novel: it crosses the $u$-cut (the cut $(c_L,c_R)$ in Fig.~\ref{fig:piNcutline}) to the adjacent Riemann sheet at tree level, disappearing from the RSII, yet remains on the complex plane of the RSII at the one-loop level.

    To further elucidate the fate of the $N^*(920)$, this work employs Baryon Chiral Perturbation Theory (B$\chi$PT) to investigate its trajectory as the pion mass increases. As a low-energy effective field theory of QCD, B$\chi$PT has been successfully applied to describe $\pi N$ elastic scattering phase shifts and the pion-nucleon $\sigma$-term. A particular advantage of B$\chi$PT in studies with unphysical pion masses is that other parameters, such as the nucleon mass $m_N$, pion decay constant $F_\pi$ and  the axial coupling constant $g_A$, can be determined self-consistently once the low-energy constants are fixed at the physical pion mass.

The paper is organized as follows. Section II gives a brief introduction to B$\chi$PT and PWAs of $\pi N$ scatterings. In  section III, the trajectory of $N^*(920)$ is presented at $O(p^2)$ and $O(p^3)$ orders for different sets of LECs values. We conclude  with a brief summary in section IV.  

\section{A brief introduction to B$\chi$PT and PWAs of $\pi N$ scatterings}
The Lagrangian in B$\chi$PT can be expanded as $
\mathcal{L} = \sum_{i=1}^{\infty} \mathcal{L}_{\pi\pi}^{(2i)} + \sum_{j=1}^{\infty} \mathcal{L}_{\pi N}^{(j)}$,
where the magnitudes of $\mathcal{L}_{\pi\pi}^{(2i)}$ and $\mathcal{L}_{\pi N}^{(j)}$ are $O(p^{2i})$ and $O(p^{j})$, respectively. Terms of the meson part for calculation up to $O(p^{4})$ are \cite{ap84gas}
\begin{align}
\mathcal{L}_{\pi\pi}^{(2)} &= \frac{F^{2}}{4} \operatorname{Tr}\left[
\nabla_{\mu}U \left(
\nabla^{\mu}U\right)^{\dagger}\right] + \frac{F^{2}}{4} \operatorname{Tr}\left[\chi U^{\dagger} + U \chi^{\dagger}\right], \label{eq:L2pi} \\
\mathcal{L}_{\pi\pi}^{(4)} &= \frac{l_{3} + l_{4}}{16} \left[\operatorname{Tr}\left(\chi U^{\dagger} + U \chi^{\dagger}\right)\right]^{2} + \frac{l_{4}}{8} \operatorname{Tr}\left[
\nabla_{\mu}U \left(
\nabla^{\mu}U\right)^{\dagger}\right] \operatorname{Tr}\left(\chi U^{\dagger} + U \chi^{\dagger}\right)~, \label{eq:L4pi}
\end{align}
where $F_\pi$ is the pion decay constant in the chiral limit. $\chi = M^{2} \mathds{1}$ (assuming isospin symmetry) and $M$ is the lowest order pion mass. Pions are contained in the SU(2) matrix:
\begin{equation}
U = \exp\left(i \frac{\phi}{F}\right), \quad \phi = \vec{\phi} \cdot \vec{\tau} = 
\begin{pmatrix}
\pi_{0} & \sqrt{2} \pi^{+} \\
\sqrt{2} \pi^{-} & -\pi_{0}
\end{pmatrix}, \label{eq:Umatrix}
\end{equation}
The covariant derivative acting on the pion fields is defined as 
$\nabla_{\mu}U = \partial_{\mu}U - i r_{\mu} U + i U l_{\mu}$,
where $l_{\mu}$ and $r_{\mu}$ are the external fields.

The required baryon Lagrangians for calculation up to $O(p^{3})$ are \cite{FETTES2000273}
\begin{align}
\mathcal{L}_{\pi N}^{(1)} &= \overline{\Psi} \left\{i 
 \slashed{D} - m + \frac{g}{2} \gamma^{\mu} \gamma_{5} u_{\mu} \right\} \Psi, \label{eq:L1N} \\
\mathcal{L}_{\pi N}^{(2)} &= \overline{\Psi} \left\{ c_{1} \operatorname{Tr}[\chi_{+}] - \frac{c_{2}}{4m^{2}} \operatorname{Tr}[u_{\mu} u_{
u}] \left(D^{\mu} D^{
u} + \text{h.c.}\right) + \frac{c_{3}}{2} \operatorname{Tr}[u^{\mu} u_{\mu}] - \frac{c_{4}}{4} \gamma^{\mu} \gamma^{
u} [u_{\mu}, u_{
u}] \right\} \Psi, \label{eq:L2N} \\
\mathcal{L}_{\pi N}^{(3)} &= \overline{\Psi} \Bigg\{ -\frac{d_{1} + d_{2}}{4m} \Big( [u_{\mu}, [D_{
u}, u^{\mu}] + [D^{\mu}, u_{
u}]] D^{
u} + \text{h.c.} \Big) 
 \\
&\quad + \frac{d_{3}}{12m^{3}} \Big( [u_{\mu}, [D_{
u}, u_{\lambda}]] \left(D^{\mu} D^{
u} D^{\lambda} + \text{sym.} \right) + \text{h.c.} \Big) + i \frac{d_{5}}{2m} \Big( [\chi_{-}, u_{\mu}] D^{\mu} + \text{h.c.} \Big) 
 \\
&\quad + i \frac{d_{14} - d_{15}}{8m} \Big( \sigma^{\mu
u} \operatorname{Tr}\left[[D_{\lambda}, u_{\mu}] u_{
u} - u_{\mu} [D_{
u}, u_{\lambda}]\right] D^{\lambda} + \text{h.c.} \Big) 
 \\
&\quad + \frac{d_{16}}{2} \gamma^{\mu} \gamma^{5} \operatorname{Tr}\left[\chi_{+}\right] u_{\mu} + \frac{i d_{18}}{2} \gamma^{\mu} \gamma^{5} \left[D_{\mu}, \chi_{-}\right] \Bigg\} \Psi, \label{eq:L3N}
\end{align}
where $m_N$ and $g$ are the bare nucleon mass and the bare axial-vector
coupling constant, respectively. Those $l_i$, $c_{i}$ and $d_{i}$ are the LECs.
The chiral vielbein and the covariant derivative with respect to the nucleon
field are defined as
\begin{align}
u_{\mu} &= i \left[ u^{\dagger} \left(\partial_{\mu} - i r_{\mu}\right) u - u \left(\partial_{\mu} - i l_{\mu}\right) u^{\dagger} \right], \\
D_{\mu} &= \partial_{\mu} + \Gamma_{\mu}, \\
\Gamma_{\mu} &= \frac{1}{2} \left[ u^{\dagger} \left(\partial_{\mu} - i r_{\mu}\right) u + u \left(\partial_{\mu} - i l_{\mu}\right) u^{\dagger} \right], \\
u &= \sqrt{U} = \exp\left(\frac{i \phi}{2F}\right). \label{eq:vielbein}
\end{align}
According to the power counting rule \cite{WEINBERG19913}, the amplitude for a diagram with $L$ loops, $I_{\phi}$ inner pion lines, $I_{N}$ inner nucleon lines and $N^{(k)}$ vertices from $O(p^{k})$ Lagrangian are of $O(p^{D})$, where 
$$D = 4L - 2I_{\phi} - I_{N} + \sum_{k}^{\infty} k N^{(k)}~.$$ 
In this manuscript, the full amplitudes of $\pi N$ scatterings are calculated up to $O(p^3)$ order.

 For the process $\pi^a(p)+N_i(q)\to \pi^{a^\prime}(p^\prime)+N_f(q^\prime)$,
 the isospin amplitude can be decomposed as: 
 \begin{equation}
T=\chi_f^{\dagger}\left(\delta^{a a^{\prime}} T^{+}+\frac{1}{2}\left[\tau^{a^{\prime}}, 
\tau^a\right] T^{-}\right) \chi_i~,
\end{equation}
where  $\tau^a$ ($a=1,2,3$) are Pauli matrices, and $\chi_i$ ($\chi_f$) corresponds to the
isospin wave function of the initial (final) nucleon state. The
amplitudes with isospins $I = 1/2, 3/2$ can be written as
\begin{equation}
\begin{aligned}
& T^{I=1 / 2}=T^{+}+2 T^{-}~, \\
& T^{I=3 / 2}=T^{+}-T^{-}~.
\end{aligned}
\end{equation}
As for Lorentz structure, for an isospin index $I=1/2,3/2$,
      \begin{equation}
T^I=\bar{u}^{(s^{\prime})}\left(q^{\prime}\right)\left[A^I(s,
t)+\frac{1}{2}\left(\slashed{p} +\slashed{p}^{\prime}\right)
B^I(s, t)\right] u^{(s)}(q),
\end{equation}
with the superscripts $(s), (s^\prime)$ denoting the spins of Dirac
spinors and  three Mandelstam variables $s=(p+q)^2, t=(p-p^\prime),
u=(p-q^\prime)$ obeying the constraint $s+t+u=2m_N^2+2m_\pi^2$. The partial wave amplitude $T^{I,J}_\pm$  for the
 $L_{2I2J}$ channel with orbital angular momentum $L$, total angular momentum  $J$ and 
total isospin  $I$ is defined as: 
\begin{equation}
   T^{I,J}_{\pm}= T(L_{2I2J})=T^{I,J}_{++}(s) \pm T^{I,J}_{+-}(s),\quad L=J\mp
    \frac{1}{2},
\end{equation}
where the definition of partial wave helicity amplitudes are written as:
\begin{equation}
\begin{aligned}
     {T}^{I,J}_{++}= & 2 m_N A^{I,J}_C(s
    )+\left(s-m_\pi^2-m_N^2\right) B^{I,J}_C(s)\\
    {T}^{I,J}_{+-}= &-\frac{1}{\sqrt{s}}\left[\left(s-m_\pi^2+m_N^2\right)
                    A^{I,J}_S(s)+m_N\left(s+m_\pi^2-m_N^2\right) B^{I,J}_S(s)\right]
\end{aligned}
\end{equation}
with
\begin{equation}\label{FSC}
    F_{C/S}^{I,J}(s)=\int_{-1}^1 \mathrm{~d} z_s F^I(s, t)\left[P_{J+1 /
2}\left(z_s\right)\pm P_{J-1 / 2}\left(z_s\right)\right],\quad F=A,B
\end{equation}
and $z_s=\cos\theta$ with  $\theta$ the scattering angle.  The partial wave
amplitudes $T^{I,J}_{\pm}$ satisfy unitarity condition:
\begin{equation}
    \operatorname{Im}T^{I,J}_{\pm}(s)
    =\rho(s,m_\pi,m_N)|T^{I,J}_{\pm}(s)|^2,\quad s>s_R=(m_\pi+m_N)^2\ .
\end{equation}
For simplicity, we denote the PWA $T(S_{11})$  as $T$ in the following.

The partial wave $S$ matrix element in $S_{11}$ channel can be defined as
\begin{equation}
    S=1+2i\rho(s)T\ ,
\end{equation}
where $\rho(s)=\sqrt{{[s-(m_N+m_\pi)^2][s-(m_N-m_\pi)^2]}}/s$.
A $K$-matrix  approximation is used to restore unitarity from perturbation amplitudes. Then, the partial wave amplitude and partial wave $S$ matrix element are expressed as
\begin{equation}
    \tilde{T}=\frac{K}{1-i\rho K},\ \tilde{S}=\frac{1+i\rho K}{1-i\rho K}\ ,
    \label{eq:tllk}
\end{equation}
where $K$ needs to be real in the physical region above the $\pi N$ threshold to meet the unitary requirement of the $S$ matrix. Usually $K$ is taken as the real part of the perturbation amplitude. For $\pi N$ scattering, it is
\begin{equation}
  \mathcal{K}^{(2)}\equiv T^{(2)}
\end{equation}
 for $O(p^2)$ calculation, while  $\mathcal{K}^{(3)} $ is set to 
 \begin{equation}
  T^{(3)}-i\rho (T^{(1)})^2
 \end{equation}
 for  $O(p^3)$ calculation, because $T^{(3)}$ contains an imaginary part on the right hand cut~\cite{Chen:2012nx}.

\begin{figure}[h!]
    \centering
    \includegraphics[width=0.4\textwidth]{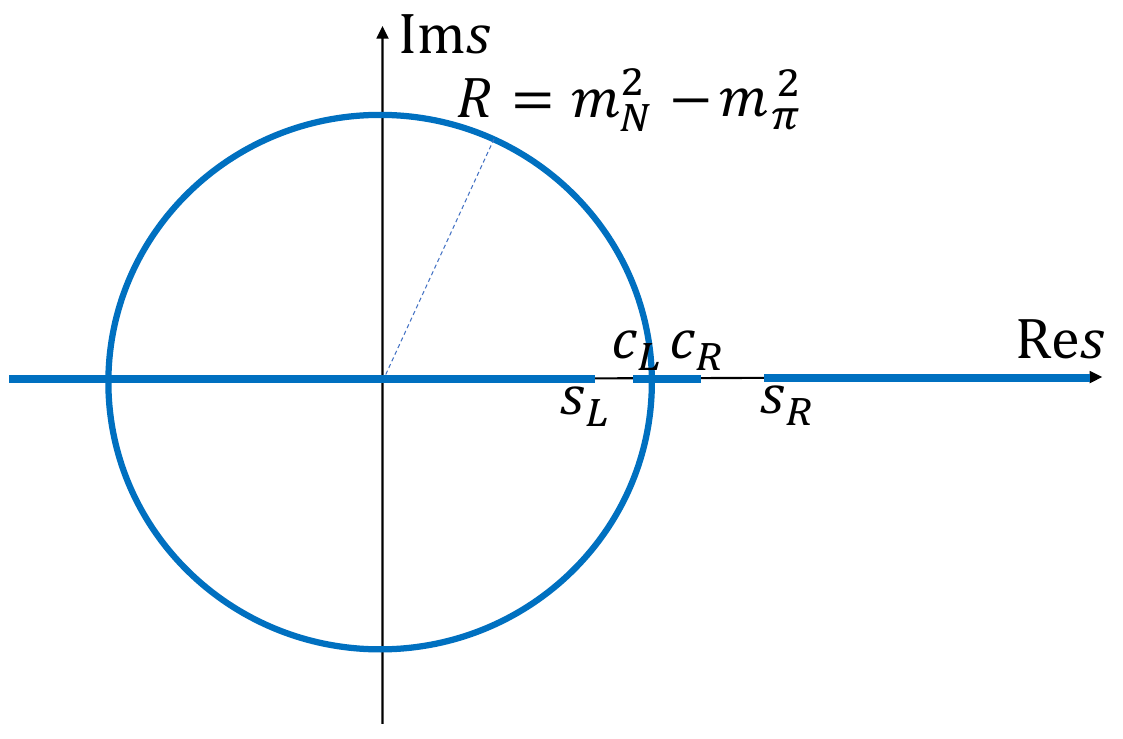}
    \caption{Cuts in $\pi N$ PWAs, represented by the bold lines. $s_L=(m_N-m_\pi)^2,c_L=(m_N^2-m_\pi^2)^2/m_N^2,c_R=m_N^2+2m_\pi^2,s_R=(m_N+m_\pi)^2$}
    \label{fig:piNcutline}
\end{figure}

The partial wave amplitude as constructed is a real analytic function on the complex $s$ plane. There exists a physical cut, or right-hand cut, above the threshold $s>(m_N+m_\pi)^2$. Partial wave projection and loop integrals also introduce other cuts, called left-hand cuts. All the cut structures in $\pi N$ scattering are shown in Fig.~\ref{fig:piNcutline}~\cite{pr59mac,Kennedy:1962ovz}. 
However, in general, such unitarization approximations suffer from problems of violation of analyticity and 
crossing symmetry~\cite{Qin:2002hk,Guo:2007ff,Guo:2007hm,Yao:2020bxx}.\footnote{For example, a [1,1] Pad\'e approximant of $\pi\pi$ scattering tend to put all contributions from different sources, e.g., s channel poles, left hand cuts, crossed channel resonance exchanges, into one single s channel resonance.} A direct consequence is the appearance of spurious physical sheet resonances (SPSRs).  
A case by case analysis seems to be required, at least, to ensure that the SPSRs play a minor contribution to physical quantities such as phase shifts. Barring for this, the K-matrix unitarization provides a quick but rough estimates of the physical pole position such as $N^*(920)$.

\section{Analysis of the $N^*(920)$ Pole Trajectory and Its Quark Mass Dependence}

To proceed, we follow Refs.~\cite{Chen:2012nx,Wang:2017agd}. First, we repeat the $O(p^2)$ and $O(p^3)$ results of  Ref.~\cite{Chen:2012nx}. 
The obtained partial wave unitary amplitude can then be used to calculate the corresponding phase shift $\delta={\rm arctan}[ \rho \tilde{T} ]$ and a subsequent fit to the phase shift data in turn  determines the  low energy constants. For the $O(p^2)$ calculations, we directly use the results in Ref.~\cite{Wang:2017agd}:
\begin{equation}
    c_1=-0.841~{\rm GeV}^{-1},\ c_2=1.170~{\rm GeV}^{-1},\ c_3=-2.618~{\rm GeV}^{-1},\ c_4=1.677~{\rm GeV}^{-1}\ .
\end{equation}
By substituting these low energy constants and physical quantities $m_N=0.9383~{\rm GeV},m_\pi=0.1396~{\rm GeV},F_\pi=0.0924~{\rm GeV},g_A=1.267$, we can calculate the cuts and poles of the partial wave unitary matrix element of the $S_{11}$ channel on the complex $s$ plane. The pole corresponding to $N^*(920)$ resonance is found  at {$\sqrt{s}=0.954\pm i0.265~{\rm GeV}$}.

In the isospin limit, the pion mass is related to the quark mass by the relation $m_\pi^2\propto 2B_0\hat{m}$, where $\hat{m}=(m_u+m_d)/2$~\cite{pr68gel}. Consequently, investigating the quark-mass dependence of the $N^*(920)$ resonance is equivalent to studying its evolution with increasing pion mass. Additionally, it is essential to determine the values of key physical quantities (e.g., $m_N$, $g_A$, and $F_\pi$) at different pion masses. Fortunately, within the framework of  B$\chi$PT, these dependence relations can be directly computed. Up to the $O(p^3)$ order (one-loop diagrams), the explicit dependence relations are given by:
\begin{equation}
    \begin{aligned}
        m_N &= m - 4 c_1 M^2 + \Delta_m,
        \quad \Delta_m  = \frac{3g^2 m_N}{32\pi^2 F^2} \left[A_0(m_N^2) + M^2 B_0(m_N^2,M^2,m_N^2)\right],\\
        F_{\pi} &= F + \Delta_F,
        \quad \Delta_F = \frac{l_4 M^2}{F} + \frac{A_0[M^2]}{16\pi^2 F},
        \quad l_4 = l_4^r + \gamma_4\lambda, \\
        \lambda &= \frac{1}{(4\pi)^2}\mu^{d-4}\left\{\frac{1}{d-4} - \frac{1}{2}(\ln 4\pi + \Gamma'(1) + 1)\right\},
        \quad l_4^r = \frac{\gamma_4}{32\pi^2}\left(\bar{l}_4 + \ln\frac{M^2}{\mu^2}\right), \quad \gamma_4 = 2,\\
        g_A &= g + 4 d_{16}M^2 + \Delta_g,\\
        \Delta_g &= \frac{g\left[4(g^2-2)m_N^2+(3g^2+2)M^2\right]}{16\pi^2 F^2(4m_N^2 - M^2)}A_0[m_N^2] + \frac{g\left[(8g^2+4)m_N^2-(4g^2+1)M^2\right]}{16\pi^2 F^2 (4m_N^2 - M^2)}\\
        &\quad +\frac{gM^2\left[-8(g^2+1)m_N^2+(3g^2+2)M^2\right]}{16\pi^2 F^2(4m_N^2-M^2)}B_0[m_N^2,m_N^2,M^2]-\frac{g^3 m_N^2(4m_N^2+3M^2)}{16\pi^2 F^2(4m_N^2-M^2)},
    \end{aligned}
    \label{eqAllM}
\end{equation}
where $A_0$ and $B_0$ denote the Passarino-Veltman functions, which are defined as~\cite{npb79pas}:
\begin{align}
    \begin{aligned}
        A_0(m^2)&=-16\pi^2 i \mu^{4-D} \int \frac{d^D k}{(2\pi)^D}\frac{1}{k^2-m^2},\\
        B_0(p^2,m_1^2,m_2^2)&=-16\pi^2 i \mu^{4-D} \int \frac{d^D k}{(2\pi)^D}\frac{1}{(k^2-m_1^2)\left[(k+p)^2-m_2^2\right]}.
    \end{aligned}
\end{align}

\begin{figure}[H]
    \centering
    \includegraphics[width=0.7\textwidth]{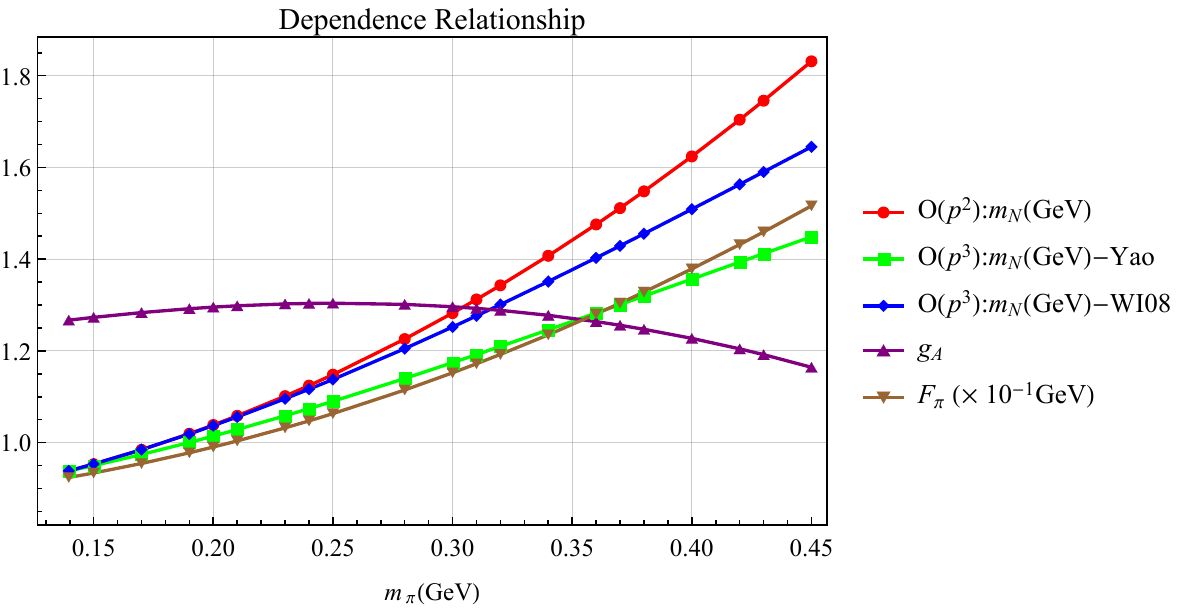} 
    \caption{Dependencies of the nucleon mass $m_N$, axial-vector coupling $g_A$, and pion decay constant $F_{\pi}$ on the pion mass $m_{\pi}$.}
    \label{fig:All(M)}
\end{figure}

Using the aforementioned formulas, the resulting dependence relations are visualized in Fig.~\ref{fig:All(M)}. The following parameter values are adopted in the calculations: $d_{16} = -0.83\,\text{GeV}^{-2}$~\cite{d16value} and $\bar{l}_4 = 4.4$~\cite{l4value}. For the three sets of $m_N$ vs. $m_\pi$ dependence curves presented in the figure, the corresponding $c_1$ parameters are chosen as follows: $c_1=-0.841\,\text{GeV}^{-1}$ for the $O(p^2)$ order~\cite{Wang:2017agd}, $c_1=-1.22\,\text{GeV}^{-1}$ for the $O(p^3)-\text{Yao}$ set~\cite{jhep16yao}, and $c_1=-1.50\,\text{GeV}^{-1}$ for the $O(p^3)-\text{WI08}$ set~\cite{WI08}.

\begin{figure}[H]
\centering
\includegraphics[width=0.8\textwidth]{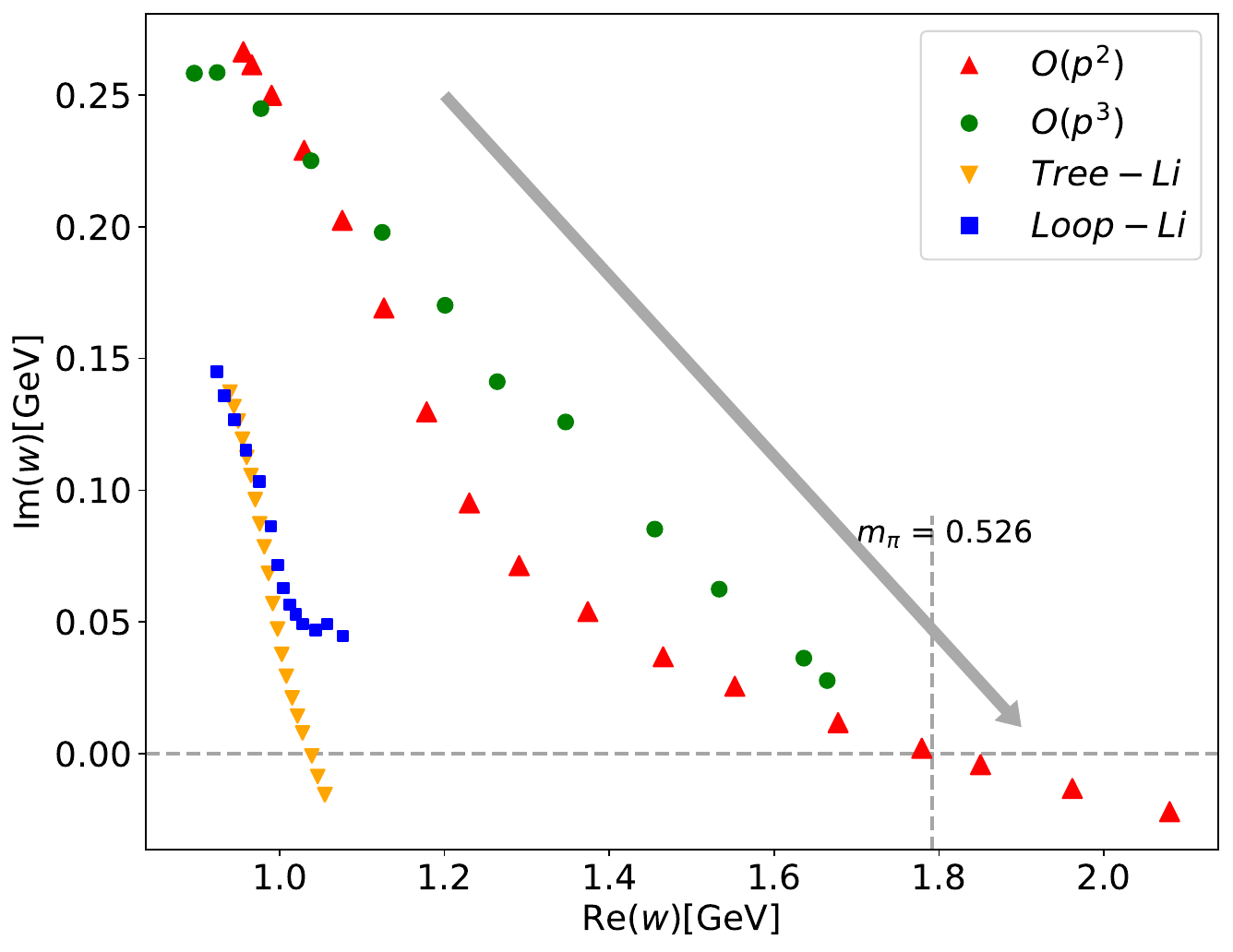}
\caption{Variation of the $N^*(920)$ pole position with the pion mass in the $\mathcal{K}^{(2)}$ and $\mathcal{K}^{(3)}$ amplitudes. The units for the pole positions are in GeV. The results obtained in this work are shown in red ($O(p^2)$ tree-level) and green ($O(p^3)$ one-loop), corresponding to pion masses in the range $m_\pi=0.1396$–$0.60\ \text{GeV}$. The results from Ref.~\cite{Li:2025fvg} are displayed in orange (Tree–Li) and blue (Loop–Li), covering the range $m_\pi=0.138$–$0.360\ \text{GeV}$.}
\label{fig:Op23w}
\end{figure}

By substituting these derived dependence relations into the partial-wave unitary matrix element, we ultimately obtain the trajectory of the $N^*(920)$ resonance as the pion mass varies from $0.1396~{\rm GeV}$ to $0.60~{\rm GeV}$. Fig.~\ref{fig:Op23w} illustrates the evolution of this $N^*(920)$ pole trajectory in the $w$-plane (where $w=\sqrt{s}$): as the pion mass increases, the pole gradually migrates toward the real axis and ultimately traverses the $u$-cut (at $m_\pi=0.526\,\text{GeV}$), thereby entering the adjacent Riemann sheet. Furthermore, the crossing position is consistent with the result calculated via Equation (43) in Ref.~\cite{Li:2025fvg}, which is expressed as:
\begin{equation}
    (m_N - m_\pi - w)(m_N + m_\pi - w)\left[m_N(m_N - w)(m_N + w)^2 - m_\pi^4\right] = 0,\quad w=\sqrt{s}
    \label{crosspoint}
\end{equation}

For the $O(p^3)$ calculations, more low energy constants are needed compared with $O(p^2)$.  We use the results of Fit 1 in Ref.~\cite{jhep16yao}(denoted as Yao in Fig.~\ref{fig:Op23w}):
\begin{equation}\label{ssecond}
    \begin{aligned}
    c_1 &= -1.22\,\text{GeV}^{-1}, \quad c_2 = 3.58\,\text{GeV}^{-1}, \quad c_3 = -6.04\,\text{GeV}^{-1}, \quad c_4 = 3.48\,\text{GeV}^{-1} \\
    d_{1}+d_{2} &= 3.25\,\text{GeV}^{-2}, \quad d_3 = -2.88\,\text{GeV}^{-2}, \quad d_5 = -0.15\,\text{GeV}^{-2} \\
    d_{14}-d_{15} &= -6.19\,\text{GeV}^{-2}, \quad d_{18} = -0.47\,\text{GeV}^{-2}
    \end{aligned}
\end{equation}
Using these low energy constants, the corresponding  positions of $N^*(920)$ pole are found to be $\sqrt{s}=0.896\pm i0.258~{\rm GeV}$. Specifically, as the pion mass increases from $0.1396~{\rm GeV}$ to $0.60~{\rm GeV}$, the trajectory of $N^*(920)$ is shown in Fig.~\ref{fig:Op23w}. Here, we have not tracked the trajectory after crossing the $u$-cut due to two reasons: first, the trajectory evolves slowly in the loop calculations; second, the error in the integral calculation becomes significant as the pole approaches the real axis. Nevertheless, we have sufficient confidence that even in the loop calculations, the $N^*(920)$ will also cross the $u$-cut and enter the adjacent Riemann sheet.

{We also compare our results with those from Ref.~\cite{Li:2025fvg}, which were obtained using the Linear Sigma Model (L$\sigma$M) with nucleons. While the overall trends of the trajectories are consistent, the rate at which the pole approaches the real axis in the L$\sigma$M is notably higher at both tree and one-loop levels. This causes the pole to cross the $u$-cut at a smaller pion mass in their tree-level calculation. Furthermore, due to the limited applicability range of the L$\sigma$M, the authors did not extend their one-loop calculation to very large pion masses. Consequently, the pole in the L$\sigma$M appears to remain on the complex plane without reaching the real axis. In contrast, here, the $O(p^3)$ result shows that the trajectory continues to bend downward toward the real axis with increasing $m_\pi$, following a trend similar to the tree-level behavior.}

In addition, we also tested another set of parameters~\cite{WI08}, referred to as WI08 parameter set, and found that the $O(p^3)$ calculation yields consistent results. The results are shown in Fig.~\ref{fignew:loop_results} below, and the specific parameters are listed as follows:
\begin{equation}\label{second}
    \begin{aligned}
    c_1 &= -1.50\,\text{GeV}^{-1}, \quad c_2 = 3.76\,\text{GeV}^{-1}, \quad c_3 = -6.63\,\text{GeV}^{-1}, \quad c_4 = 3.68\,\text{GeV}^{-1} \\
    d_{1}+d_2 &= 3.67\,\text{GeV}^{-2}, \quad d_3 = -2.63\,\text{GeV}^{-2}, \quad d_5 = -0.07\,\text{GeV}^{-2} \\
    d_{14}-d_{15} &= -6.80\,\text{GeV}^{-2}, \quad d_{18} = -0.50\,\text{GeV}^{-2}\,.
    \end{aligned}
\end{equation}

\begin{figure}[H]
\centering
\includegraphics[width=0.8\textwidth]{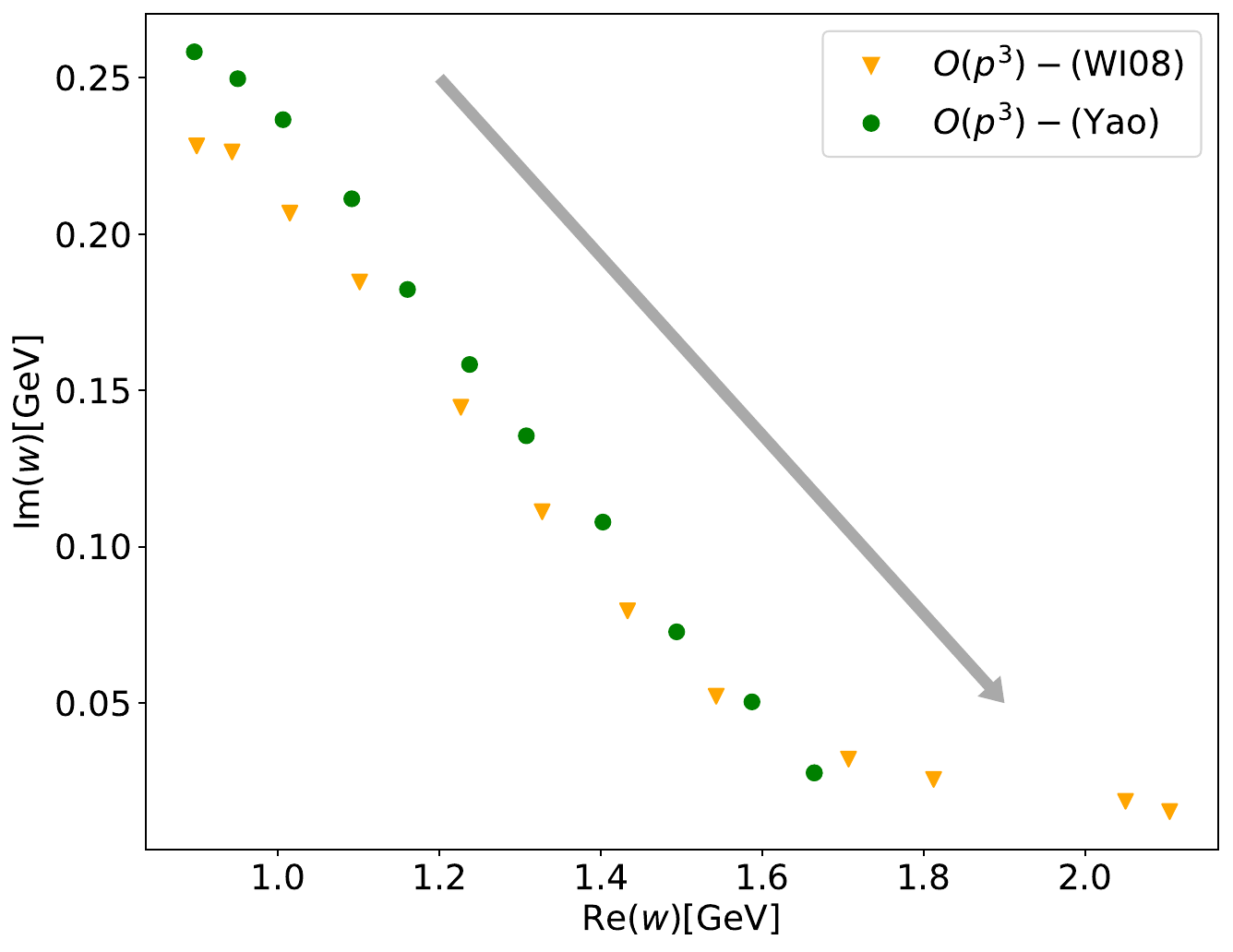}
\caption{$m_\pi$ dependence of $N^*(920)$ pole from the full $O(p^3)$ amplitude including loop corrections, using parameters from Eq.~(\ref{second}).}
\label{fignew:loop_results}
\end{figure}

\begin{figure}[H]
    \centering
   \includegraphics[scale=0.5]{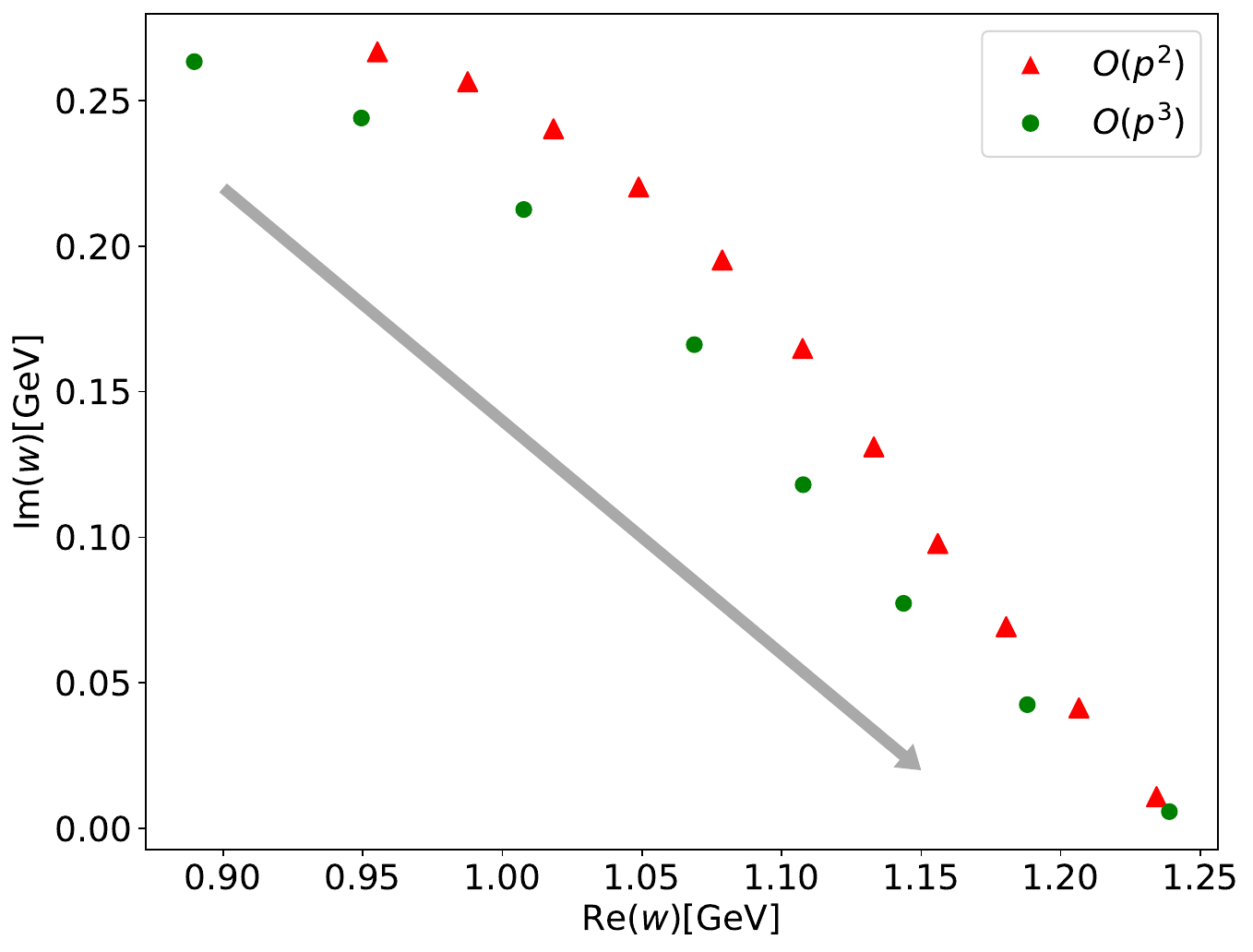}
    \caption{The dependence of the $N^*(920)$ pole position on pion mass, as determined from the $\mathcal{K}^{(2)}$ and $\mathcal{K}^{(3)}$ amplitudes (with the dependencies of $m_N$, $g_A$, and $F_\pi$ on $m_\pi$ taken from lattice–data–based fits). The unit is ${\rm GeV}$. The $\mathcal{K}^{(2)}$ results are indicated by red triangles, while the $\mathcal{K}^{(3)}$ results are shown as green circles. The pion mass $m_\pi$ varies from 0.1396 to 0.44 GeV.}
    \label{fig:poletracelattice}
\end{figure}

In addition to the dependence relations for $m_N$, $g_A$ and $F_\pi$ derived from chiral perturbation theory, similar results are also available by some theoretical fits performed on the lattice data. For $m_N$, we use the ruler approximation in Ref.~\cite{pos14wal}, that is, $m_N=800~{\rm MeV}+m_\pi$, which is consistent with the lattice QCD {results~\cite{prd13gon}} in a large range. For $g_A$, we use the $O(p^3)$ result in Ref.~\cite{prd22alv}, and for $F_\pi$, we use the fit result with strategy 2 in Ref.~\cite{prl21nie}. 
Based on these dependence relations, the resulting trajectory of the $N^{*}(920)$ pole is shown in Fig.~\ref{fig:poletracelattice}. It seems that the points where the $N^{*}(920)$ pole approaches the $u$-cut from the $O(p^2)$ and $O(p^3)$ chiral perturbation theory calculations tend to converge. However, since there is no guarantee that the pole reaches the $u$-cut at the same $m_\pi$ for both $O(p^2)$ and $O(p^3)$ results, this convergence may just be  accidental.

As illustrated in this section, the $N^*(920)$ pole trajectory  obtained in different approximations and parameters are in qualitative agreement with each other.

\section{Summary}\label{sec:s}

In this paper, we have investigated the trajectory of $N^*(920)$ as the pion mass increases within the B$\chi$PT framework both  at $O(p^2)$ and $O(p^3)$ orders. In B$\chi$PT, the functions of the nucleon mass, pion decay constant, and $\pi N$ axial-vector coupling as a function of the pion mass are obtained self-consistently, provided that the LECs are fixed. In both cases, the $N^*(920)$ moves along a rightward-downward trajectory toward the $u$-cut on the complex energy plane, and may eventually cross the $u$-cut, entering the adjacent Riemann sheet defined by the $u$-cut. The result at $O(p^3)$ order shows that the circular cut has marginal effects on the trajectory, and the higher-order contributions only slightly alter the approaching rate of the pole. Furthermore, to test the robustness, we fix the LECs at three different parameter sets. Consequently, all three results demonstrate that the $N^*(920)$ moves toward the $u$-cut and eventually enters the adjacent Riemann sheet. The trajectory is also compared with that in the previous work~\cite{Li:2025fvg}, showing qualitatively consistent behaviors. Our analyses made in this paper may provide valuable   insights for future Lattice studies with unphysical pion masses.

\vspace{0.2cm}
{\bf Acknowledgment:} This work is supported by China National Natural Science Foundation under Contract No. 12335002, 12375078.
This work is also supported by “the Fundamental Research Funds for the Central Universities”.

\newpage

\bibliographystyle{utphys}
\bibliography{inspire_cite}

\appendix

\end{document}